\documentclass[11pt, oneside]{article} 	
\usepackage[margin=1in]{geometry} 		
\usepackage[parfill]{parskip} 		
\usepackage{graphicx}				
\usepackage{amssymb,natbib}

\usepackage{amssymb,amsmath,bm,microtype,setspace,natbib,subcaption,upgreek}
\usepackage[colorlinks=true,linkcolor=blue,citecolor=blue]{hyperref}

\usepackage{tikz}
\usetikzlibrary{positioning, arrows.meta}

\DeclareMathOperator*{\argmax}{arg\,max}

\title{Some Tradeoffs of Competition in Grant Contests}
\author{Kyle R. Myers\thanks{Harvard Business School, \href{mailto:kmyers@hbs.edu}{kmyers@hbs.edu}. This paper would not have been possible without Kevin Gross sharing both code and thoughtful comments. I am also grateful for feedback from Heidi Williams, Carl Bergstrom, Charles Ayoubi, Manuel Hoffman, Pierre Azoulay, and Wei Yang Tham. All errors are my own.}}
\date{March 1, 2024}

\begin{document}
\maketitle
\begin{abstract}
When funding public goods, resources are often allocated via mechanisms that resemble contests, especially in the case of research grants. A common critique of these contests is that they induce ``too much'' effort from participants. This need not be true if the effort in the contest is itself directed towards the public good. This papers analyzes survey data on scientists' time use and finds that scientists allocate their time in a way that is consistent with fundraising effort (e.g., grant writing) having inherent scientific value---scientists who spend more time fundraising do not spent significantly less time on research even after conditioning on confounding factors. Theoretical models of contests are used to show that the presence of such a positive effort externality, where scientists generate social value when pursuing grants, changes the relationship between competition and the aggregate productivity of a grant contest. Ensuring that scientists exert socially valuable effort to obtain grants is increasingly important as grant contests become more competitive.
\-\\
\-\\ \emph{Keywords}: contests; public goods; research grants; scientific funding; competition
\end{abstract}

\clearpage
\onehalfspacing
\section{Introduction}

There is no shortage of critiques claiming that competition for research grants incentivizes scientists to spend ``too much'' time applying for funding. Furthermore, some critics have suggested that funders should introduce randomized lotteries or reduce application requirements.\footnote{For example, see: \cite{herbert2013time,fang2016research,gallup2016hijacked,vaesen2017much,gross2019contest,piper2019science,else2021covid,santoro2021daunting,dresler2022many,ross2022competitive,thompson2022silicon,smith2024spending}.} More generally, it has long been appreciated that competitions structured as contests---where participants sink effort in exchange for increasing the probability they win a prize---can induce over-investment relative to the social optimum when participants' efforts are unproductive (\citealt{loury1979market}).

\begin{figure}[htbp]\centering
\caption{Competition at science funding agencies}
\label{fig_competition} 
\subfloat[Across major funders]{
\includegraphics[width=0.475\linewidth, trim = 0mm 0mm 0mm 0mm , clip]{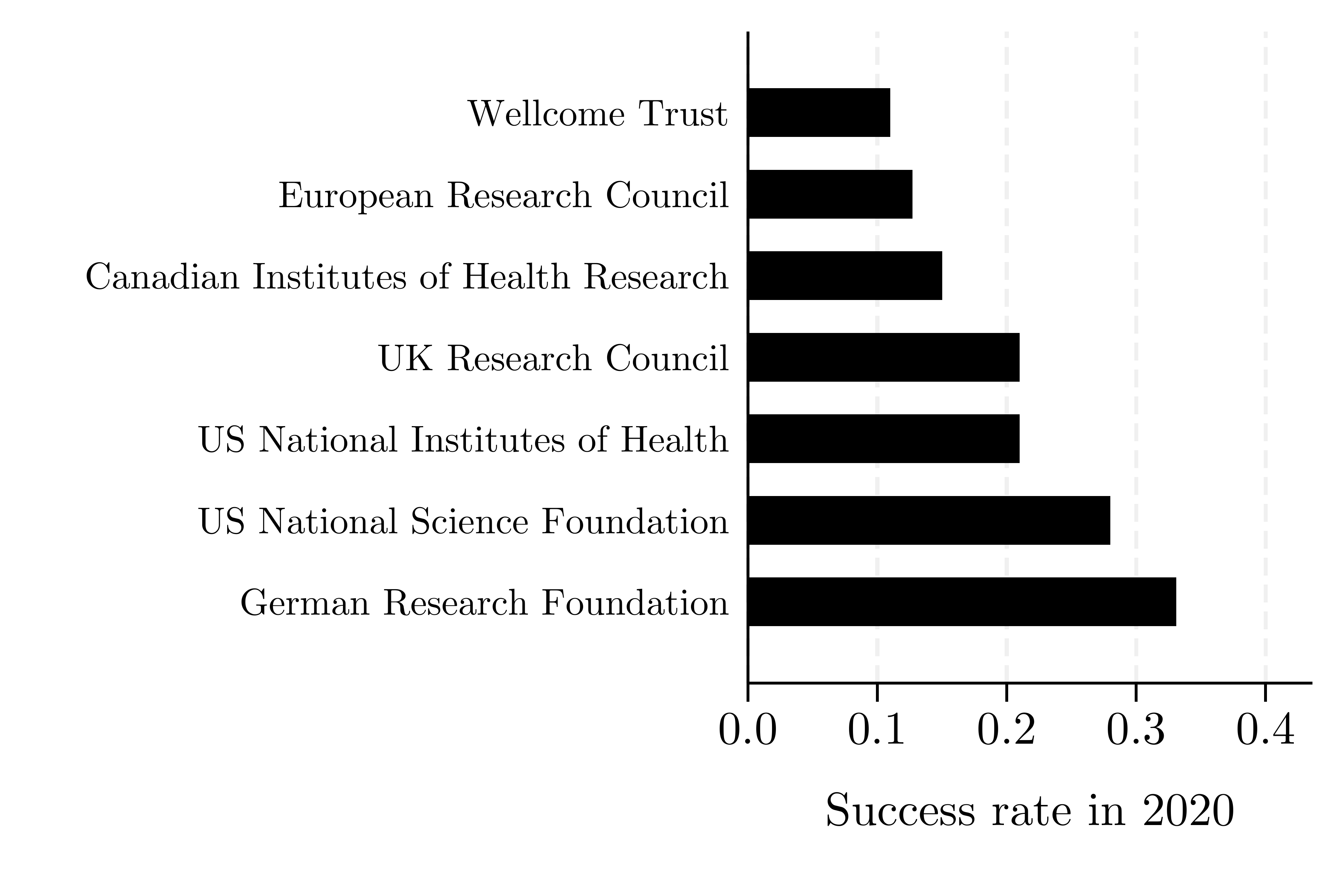}}
\subfloat[US National Institutes of Health over time]{
\includegraphics[width=0.475\linewidth, trim = 0mm 0mm 0mm 0mm , clip]{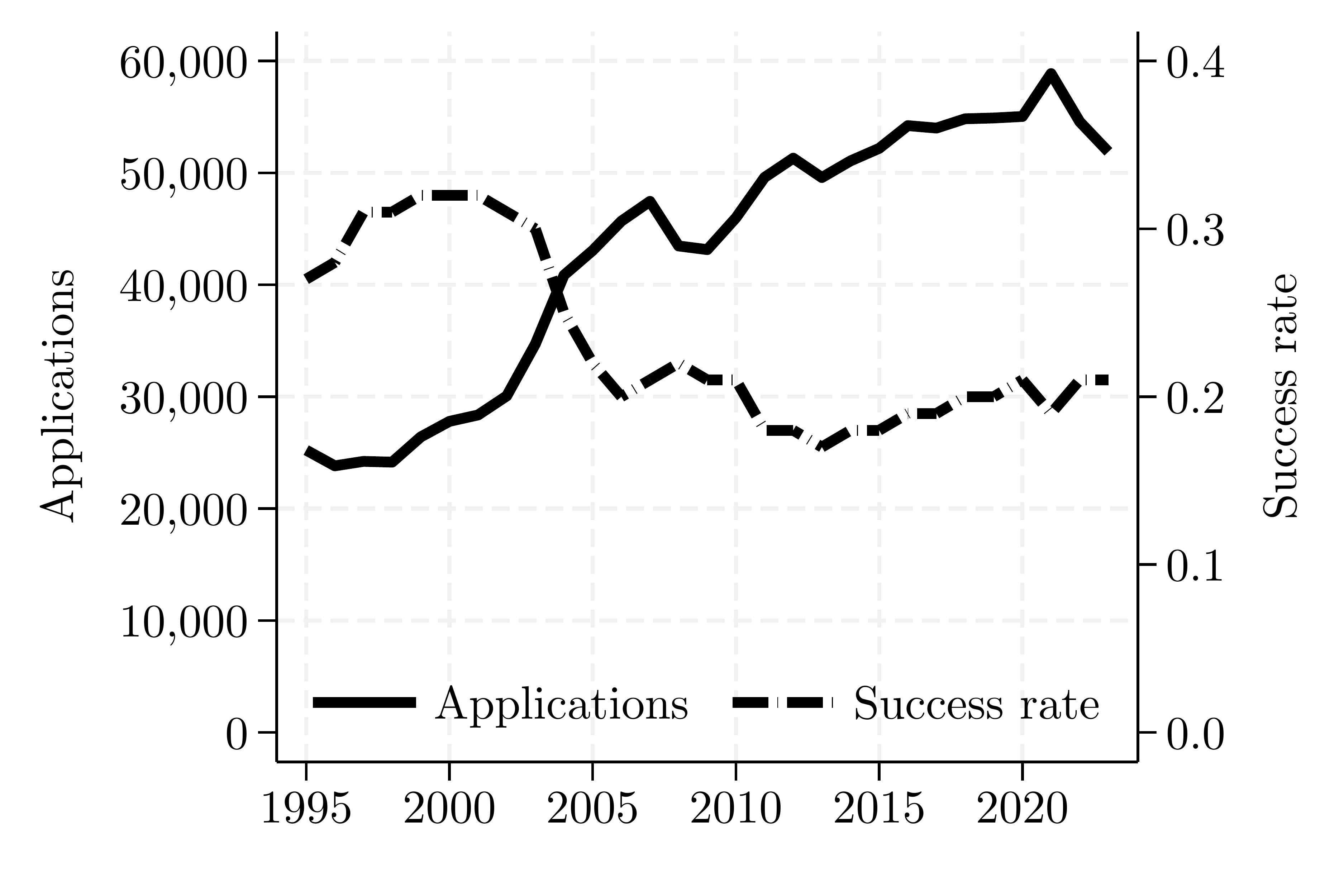}}
\begin{quote}\scriptsize
Note: Panel (a) shows the success rates for major funders in 2020. Panel (b) shows the trend in the number of annual applications and success rates for ``Research Project Grants'' at the US National Institutes of Health.
\end{quote}
\end{figure}

Obtaining a research grant is certainly a difficult and competitive endeavor. Using success rates as a proxy for competition, Figure \ref{fig_competition} shows that scientists pursuing a grant from a range of major funders currently have a 10--30\% chance of success with significant variation across funders (Panel a) and within funders over time (Panel b).\footnote{Data for funders shown in Figure \ref{fig_competition} are as follows: Wellcome Trust (applications: 3,584; success rate: 0.11) European Research Council (applications: 9,304; success rate: 0.13) Canadian Institutes of Health Research (applications: 4,488; success rate: 0.15) US National Institutes of Health (applications: 55,038; success rate: 0.21) UK Research Council (applications: 19,497; success rate: 0.21) US National Science Foundation (applications: 42,726; success rate: 0.28) German Research Foundation (applications: 14,700; success rate: 0.33)
 (\citealt{wellcome2021grant,erc2024dashboard,cihr2021statistics,ukri2021competitive,nih2023data,nsf2022funding,dfg2020facts}).} This  competition requires scientists to spend a significant amount of time on their fundraising efforts. In the sample I study in this paper (described below), scientists spend an average of roughly 7 hours per week (s.d. 4.8; 15\% of their time) developing, writing, and submitting grant applications. But how should we judge this variation; are more competitive funding environments inefficient; is fundraising effort a waste; or are there important tradeoffs to consider?

\clearpage

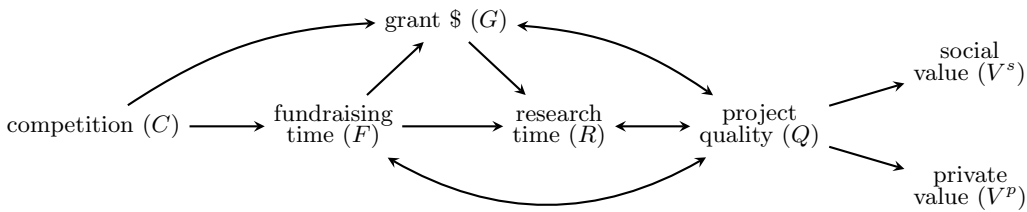
\begin{figure}[htbp]\centering\footnotesize
\caption{Conceptual framework}
\label{fig_framework}
\vspace{1em}
\begin{tikzpicture}[->, >=stealth, node distance=1.5cm and 1cm, thick, main node/.style={draw, rectangle, align=center, inner sep=4pt},open node/.style={align=center}]
	\node[open node] (F) {fundraising \\[-0.3em] time ($F$)};
	\node[open node] (C) [left = 1cm of F] {competition ($C$)};
	\node[open node] (empt1) [right of = F] {};
	\node[open node] (R) [right of = empt1] {research \\[-0.3em] time ($R$)};
	\node[open node] (G) [above = 1cm of empt1] {grant \$ ($G$)};
	\node[open node] (Q) [right = 1cm of R] {project \\[-0.3em] quality ($Q$)};
	\node[open node] (Vs) [above right = 0cm and 1cm of Q] {social \\[-0.3em] value ($V^s$)};
	\node[open node] (Vp) [below right = 0cm and 1cm  of Q] {private \\[-0.3em] value ($V^p$)};
	\draw (C) -- (F);
	\draw (C) to [bend left=15] (G) ;
	\draw (F) -- (G);
	\draw (F) -- (R);
	\draw (G) -- (R); 
	\draw[<->] (R) -- (Q);
	\draw[<->] (F) to [bend right=30] (Q) ;
	\draw[<->] (Q) to [bend right=15] (G) ;
	\draw (Q) -- (Vs);
	\draw (Q) -- (Vp);
\end{tikzpicture}\vspace{1em}
\begin{quote}\scriptsize
Note: Shows a directed graph of the key elements that determine how the level of competition a scientist faces for grant funding can ultimately affect their scientific output.
\end{quote}
\end{figure}

Figure \ref{fig_framework} provides a conceptual framework to organize (some of) the tradeoffs here. It is drawn from the perspective of an individual scientist and illustrates how competition they face for funding ($C$) can potentially affect their time allocations ($F$ and $R$), the amount of grant funding they obtain ($G$), the quality of their ideas ($Q$), the value created by those ideas that accrues to the scientist ($V^p$), and the total value to society ($V^s$).

A few points are worth emphasizing here. First, a key point of this framework is that the quality of ideas scientists are working on will affect their time allocations and the amount of resources they can obtain (e.g., a scientist with a great idea may spend more time fundraising, obtain more funding, and spend more time on research as well).\footnote{Furthermore, note that quality could reflect notions of speed (e.g., the same project completed sooner is more valuable to society).} Second, it has long been appreciated that scientists capture a tiny fracture of the social value they generate (\citealt{nordhaus2004schumpeterian,lakdawalla2010economic}), so improvements in quality are much more valuable to society than they are the scientist ($V^s>>V^p$). Third, in this framework, and throughout this paper, I will use the terms ``fundraising'' and ``research'' to delineate effort that is best classified by those labels; however, as indicated by the path between $F$ and $Q$ in Figure \ref{fig_framework} and supported by evidence discussed below, ``fundraising'' may very well include scientific efforts that are socially valuable.\footnote{This framework treats each component as a single, scalar variable and abstracts away from potentially important multi-dimensional issues where, for example, fundraising effort improves quality only on dimensions related to the private value (to the scientist) of the project. Furthermore, it abstracts away from potentially important sources of heterogeneity as well as the possibility that competition might influence the allocation process (e.g., the cost to a funder of identifying the most promising ideas may be a function of competition for the funder's funding). I return to these and other limitations in the Discussion section.} Lastly, this framework abstracts away from the mechanisms of competition completely. ``Competition'' could reflect more competitors competing for the same grants, it could reflect the amount of preliminary results required to receive a grant, etc.\footnote{For more on competition for grants in the context of relative ranking mechanisms, see \cite{adda2023grantmaking}.}

It is useful to phrase the aforementioned critiques through the lens of this framework. In short, some critics claim that the amount of fundraising scientists undertake detracts from their research ($F \rightarrow R$ is negative) and/or generates no value ($F \rightarrow Q$ is non-positive); it is a waste. The specific mechanism underlying these claims varies, but often can be traced to competitive pressures.\footnote{Obviously, any activity that does not either (a) improve a funders' ability to efficiently allocate their funding or (b) improve the social value of a scientists' idea is wasteful by definition.} This motivates proposals for mechanisms that dampen the returns to fundraising effort (decrease $F \rightarrow G$) such as lotteries or applications that do not require preliminary results.

How plausible are these claims about wasted efforts? All scientists face time constraints. So, holding fixed the amount of funding a scientist could expect to obtain (conditional on $G$), more fundraising could decrease time spent on research. But that fundraising effort may serve two purposes. First, it may be necessary to identify scientists with high-quality ideas and therefore is a source of allocative efficiency (i.e., how well the contest can identify scientists with higher quality ideas). Second, fundraising effort may itself be inherently valuable if it improves the quality of the ideas submitted and therefore is a source of aggregate productivity (i.e., the total benefits less costs). Second, it may be inherently valuable as a determinant 

This paper is focused on the latter possibility---the inherent value of fundraising efforts. If fundraising effort is itself important, it could imply two features: (1) the net effect of fundraising on research conditional on grant funding (per $F \rightarrow R$ and $F \rightarrow Q \rightarrow R$) need not be negative; and (2) the net effect of competition on social value ($C \rightarrow ... \rightarrow V^s$) need not be negative.

To date, surprisingly little empirical work has investigated these relationships. It is unclear how shifts in fundraising time affect shifts in research time. But there is some evidence that we should not ignore the possibility that fundraising effort may improve project quality and spark more research time. First, applications almost always require some degree of ``science'' to be performed. As evidence to this, scientists participating in federal grant competitions report that approximately 38\% of their time spent on proposal preparation contributes directly to their scholarship (\citealt{schneider2018results}).\footnote{It seems reasonable to treat this self-reported estimate as a lower bound since it would be in scientists' self-interest to report a low estimate on a survey focused on the (private) costs of grant applications.} Furthermore, studies of science funding agencies typically find that (1) scientists who compete for these grants publish much more than those who do not apply, (2) applicants tend to publish ideas related to their applications whether they are funded or not, and (3) the estimated treatment effect of receiving these grants based on comparisons of the publication output of funded and un-funded applicants appears relatively small or even negative (\citealt{jacob2011impact,li2017expertise,wang2019early,ayoubi2019important,myers2020elasticity}). These patterns could be described by selection effects dominating the returns to the grant funding, but they are also consistent with these competitions incentivizing significant amount of productive scientific effort.

In this paper, I provide new evidence on scientists' time allocations and the relationship between fundraising and research efforts. First, I analyze survey data from a sample of research-active professors at major US universities. In the survey, professors report their expected time use and funding over the coming five years, which I use to uncover how they view the relationship between fundraising and research per their own stated time allocations and funding expectations. In order to estimate the net effect of fundraising on research conditional on grant funding (i.e., the net effect of $F \rightarrow R$ and $F \rightarrow Q \rightarrow R$ conditional on the effect of $F \rightarrow G \rightarrow R$), I use a large set of control variables as well as an instrumental variable in an attempt to isolate variation in fundraising that is driven by competition. Although the data and my empirical approach are not without significant limitations, which I discuss throughout the paper, they provide a novel, quantitative view of relationships that have largely been debated only via anecdote and opinion. Overall, I cannot reject a null effect---conditional on funding, scientists who plan to spend more time on fundraising do not expect to spend any less (or more) time on research. Notably, I do estimate that the effect of fundraising on research to be significantly smaller (in absolute terms) than the effect of scientists' other duties (e.g., teaching, administration) on their research---if fundraising time does reduce research time, it does so at a significantly lower rate than other demands on professors' time.

The opportunity cost of fundraising time appears to be significantly lower than the opportunity cost of other duties. The results are consistent with the following model: the negative effects of fundraising on research driven by time constraints are offset by the positive effects driven by the improvement in a project's quality that spur more research effort. This is socially efficient if the social value generated by this quality improvement is greater than the cost to the scientist of this additional fundraising effort. The positive externalities of science are thought to be very large (i.e., $\frac{V^p}{V^s}$ is very small); so, this seems plausible.

In order to better understand how the productivity of a grant contest depends on the extent to which fundraising effort generating social value, the second analyses in this paper uses theory to illustrate the importance of effort externalities in contests. Using both an elementary and a more realistic model (building closely off of \citealt{gross2019contest}), I show the (social) benefit of inducing (privately) costly effort in contests for funding public goods, so long as that work is directed towards the public good, here, ``science.'' This echoes results in other work on how the presence of positive externalities can lead contests to have desirable social properties (e.g., \citealt{morgan2000financing}). The rest of the paper is as follows: Section \ref{sec_empirics} contains the empirical analyses of the survey data; Section \ref{sec_theory} contains the theoretical analyses of effort externalities in contests; and Section \ref{sec_discuss} concludes with a discussion of the limitations and implications of this work.

\section{Empirics: Does fundraising spur or steal research?}\label{sec_empirics}
\subsection{Research design}

Ideally, all of the components of Figure \ref{fig_framework} would be observable and I could trace an exogenous source of variation in competition through the model to see how it affects each component. That is not currently possible. However, using the survey data described below, I can make some progress on understanding a portion of the relationships.

\begin{figure}[htbp]\centering\footnotesize
\caption{Empirical model}
\label{fig_empmodel}
\begin{tikzpicture}[->, >=stealth, node distance=2cm and 2cm, thick, main node/.style={draw, rectangle, align=center, inner sep=4pt}, open node/.style={align=center}]
	\node[main node] (F) {fundraising \\[-0.3em] time ($F$)};
	\node[open node] (empt1) [right of = F] {};
	\node[main node] (R) [right of = empt1] {research \\[-0.3em] time ($R$)};
	\node[main node] (G) [above = 1cm of empt1] {grant \$ ($G$)};
	\node[open node] (omega) [above = 0.7cm of G] {$\omega$};
	\node[open node] (Z) [above left = 1cm and -0.5cm of F] {$Z$};
	\node[open node] (empt2) [below right = 0.2cm and 0.2cm of G] {}; 
	\draw (F) -- (G) ;
	\draw (F) -- (R) node[midway, below] {$\beta_{F}$};
	\draw (G) -- (R); 
	\draw[dotted] (omega) to [bend right=40] (F);
	\draw[dotted] (omega) -- (G);
	\draw[dotted] (omega) to [bend left=40] (R);
	\draw (Z) -- (F);
	\draw[dotted] (Z) -- (G);
\end{tikzpicture}\vspace{1em}
\begin{quote}\scriptsize
Note: Shows a directed graph of the empirical analyses where variables in boxes are observable data, $\omega$ represents an unobservable shock that jointly affects scientists' time allocations and grant funding (e.g., a promising new idea), and $Z$ represents either an unobservable source of variation or an instrumental variable.
\end{quote}
\end{figure}
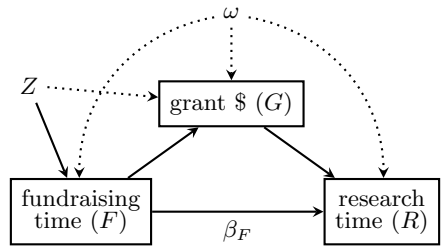

To organize my empirical approach, Figure \ref{fig_empmodel} presents a reduced-form representation of the conceptual framework that focuses on the variables I can observe and the relationships I can estimate. Since ``quality'' is notoriously unobservable, I cannot separately estimate the different paths through which fundraising time could affect research time. Thus, I focus on estimating $\beta_F$, which reflects the effect of fundraising time on research time, conditional on the grant dollars generated by the fundraising. Referring back to Figure \ref{fig_framework}, $\beta_F$ is the net effect of $F \rightarrow R$ and $F \rightarrow Q \rightarrow R$ (e.g., the net effect of any substitution away from research due to time constraints versus any complementarities driven by an improvement in quality). This is a policy-relevant parameter that is necessary for evaluating policies that shift competition in research grant contets.\footnote{The indirect effect of fundraising time on grant funding and, in turn, the effect of that funding on research time are also important, policy relevant relationships to investigate. However, as will be clear below, I lack a source of variation in fundraising time that is not also plausibly correlated with grant funding itself. This prevents me from identifying $F \rightarrow G$ and $G \rightarrow R$. The returns to research grant funding ($G \rightarrow R$) is a popular, but still very active topic for obvious reasons (See: \citealt{jacob2011impact,li2017expertise,wang2019early,ayoubi2019important,myers2020elasticity}).}

The key challenge when using non-experimental data to estimate $\beta_F$, as I will attempt to do, is ensuring that there is no unobservable factor jointly determining the outcomes. Figure \ref{fig_empmodel} shows this possibility with $\omega$, which is analogous to the quality factor ($Q$) in the framework of Figure \ref{fig_framework}. The concern is, for example, that a scientist who has a high-quality idea they want to pursue may (i) spend more time fundraising, (ii) receive more funding per time spent fundraising, and (iii) spend more time on research.

I lack a silver bullet to handle this sort of endogeneity concern, but I take two steps to address it as much as possible. First, I use Lasso to select from a high-dimensional set of possible controls in an attempt to reduce the possibility that an endogenous factor such as $\omega$ is influencing the (residual) variation. If I can eliminate all variation driven by $\omega$ then any variation left in fundraising time should be driven by some other force ($Z$) that affects just fundraising time (and possibly grant funding, $G$) but not research time (e.g., the amount of competition a scientist expects when fundraising). I can't ever guarantee that the residual variation does not affect $G$. But, since my focus is $\beta_F$, I'll always be conditioning on the mediator $G$.\footnote{Since this mediator $G$ is not as-good-as randomly assigned, I do not report nor interpret the relationships between $G$ and $R$.}

Instead of relying on residual variation (i.e., leaving $Z$ unobserved), my second approach is to identify a proxy for competition in the data and use it as an instrumental variable that satisfies the necessary assumptions.\footnote{Those assumptions are relevance ($Z \rightarrow F$), independence ($Z \not\leftrightarrow \omega$), and exclusion ($Z \not\leftrightarrow R$). I will always be conditioning $Z$ on $G$ in order to shut down that indirect causal path from $Z$ to $R$} I explore a number of alternative proxies for competition, but my primary measure is based on field-level variation in how much funding scientists expect to obtain per hour of fundraising time: $G/F$. Intuitively, if all scientists in a field report that they expect to obtain fewer grant dollars per fundraising hour, they are more likely to be in a more competitive environment and less likely to spend time fundraising on average. In order to generate this proxy in a way that is not biased by each individuals' data point I construct a jackknife or ``leave-one-out'' average of $G/F$, where the average value assigned to each professor is the average of $G/F$ in their field calculated without their own value.\footnote{Specifically, for scientist $i$ in field $d$ where $N_d$ is the number of scientists in the field: $Z_{id} \equiv \frac{1}{N_d-1} \sum_{i'=1}^{N_d} \frac{G_{i'}}{F_{i'}} \,\forall i' \neq i$.}

\subsection{Data}

All data here comes from \cite{myers2023new}. It is a survey of professors at major US institutions of higher education across all fields of science. See  \cite{myers2023new} for full description of the population definition, sampling methodology, and survey instrument. In total, 131,672 professors at roughly 150 universities were invited and 4,388 (3.33\%) completed the survey.

\begin{table}[htbp]\centering\footnotesize
\caption{Summary statistics}
\label{tab_sumstat}
{
\def\sym#1{\ifmmode^{#1}\else\(^{#1}\)\fi}
\begin{tabular}{l*{1}{rrrrr}}
\hline\hline
                    &       count&        mean&          sd\\
\hline
\underline{\emph{Broad field} \{0,1\}}&            &            &            \\
Engineering, math, \& related&       2,640&        0.20&        0.40\\
Humanities \& related&       2,640&        0.16&        0.37\\
Medical \& health sciences&       2,640&        0.27&        0.44\\
Natural sciences    &       2,640&        0.20&        0.40\\
Social sciences     &       2,640&        0.17&        0.38\\
[0.5em] \underline{\emph{Professor rank \& tenure} \{0,1\}}&            &            &            \\
Assistant           &       2,640&        0.29&        0.45\\
Associate           &       2,640&        0.25&        0.44\\
Full                &       2,640&        0.41&        0.49\\
Other               &       2,640&        0.05&        0.21\\
Tenure track        &       2,640&        0.27&        0.44\\
Tenured             &       2,640&        0.59&        0.49\\
Non-tenure track    &       2,640&        0.14&        0.34\\
[0.5em] \underline{\emph{Time use, hrs. per week} (0,$\infty$)}&            &            &            \\
Research            &       2,640&       19.53&        8.80\\
Fundraising         &       2,640&        6.95&        4.84\\
Teaching            &       2,640&       12.45&        7.45\\
Administration, clinical, etc.&       2,640&       11.81&        8.75\\
All non-research/fundraising&       2,640&       24.15&       10.75\\
[0.5em] \underline{\emph{Funding, \$-M per year} [0,$\infty$)}&            &            &            \\
Fundraising expectations&       2,640&        0.13&        0.16\\
Guaranteed funding  &       2,640&        0.08&        0.12\\
\hline\hline
\end{tabular}
}

\vspace{0.5em}
\begin{quote}\scriptsize
Note: Reports the mean and standard deviation (sd) of key variables used in the analyses. \{0,1\} indicates the variable is a binary indicator and the other interval notation indicates the range of possible support.
\end{quote}
\end{table}

Starting with this full sample, I first restrict the sample to professors reporting non-zero research time (4,186; 95.40\%). Then, I restrict the sample to those reporting non-zero time on other work including teaching or administration (4,135; 98.78\%). Lastly, I restrict the sample to those reporting non-zero fundraising time (2,640; 63.85\%).
 The first restriction ensures all professors are active scientists. The second is useful because I use professors' time on these other activities as a benchmark of time-constraint driven substitution across tasks.\footnote{It also ensures the sample doesn't include professors in particularly unique circumstances where all they do is research.} The last restriction significantly reduces the sample size; however, it ensures that all professors in the analyses are at risk of fundraising effort. Furthermore, it implies that the results are driven by (and therefore relevant to) changes in fundraising effort on the intensive margin.

The top of Table \ref{tab_sumstat} reports the share of respondents from five broad field groupings as well as some other descriptors related to the professors' rank and tenure status. In the Appendix, Table \ref{tab_sumstat_dis} provides the share of professors in each of twenty narrower fields. \cite{myers2023new} provides a much more detailed view of the sample. Including the high-dimensional set of potential control variables I include in some regressions.

\begin{figure}[htbp]\centering
\caption{Distributions and correlations of time use}
\label{fig_sumstat} 
\subfloat[Average time allocation]{
\includegraphics[width=0.85\linewidth, trim = 0mm 0mm 0mm 10mm , clip]{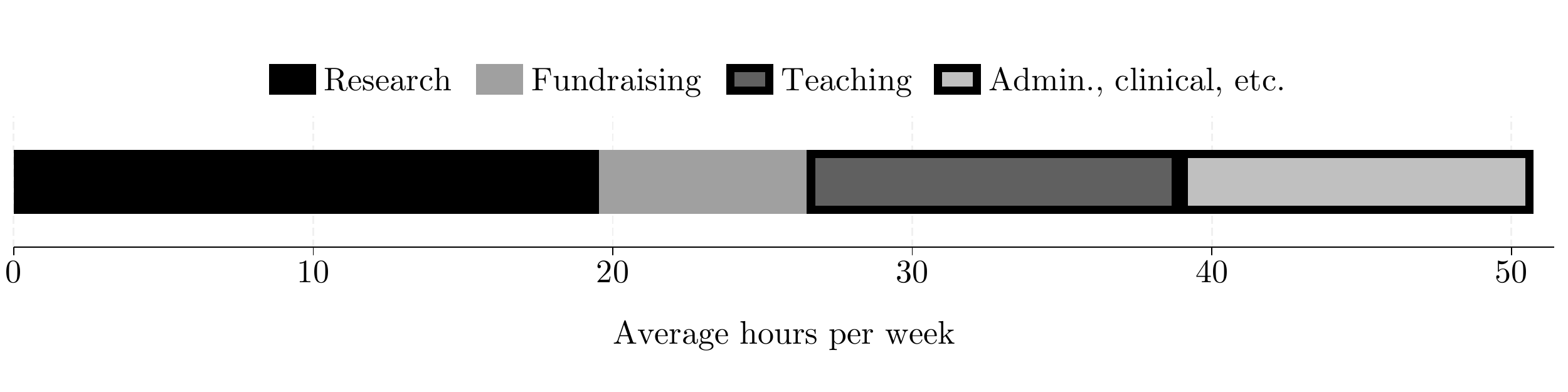}} \\ \-\\ 
\subfloat[Distributions of time]{
\includegraphics[width=0.475\linewidth, trim = 0mm 0mm 0mm 0mm , clip]{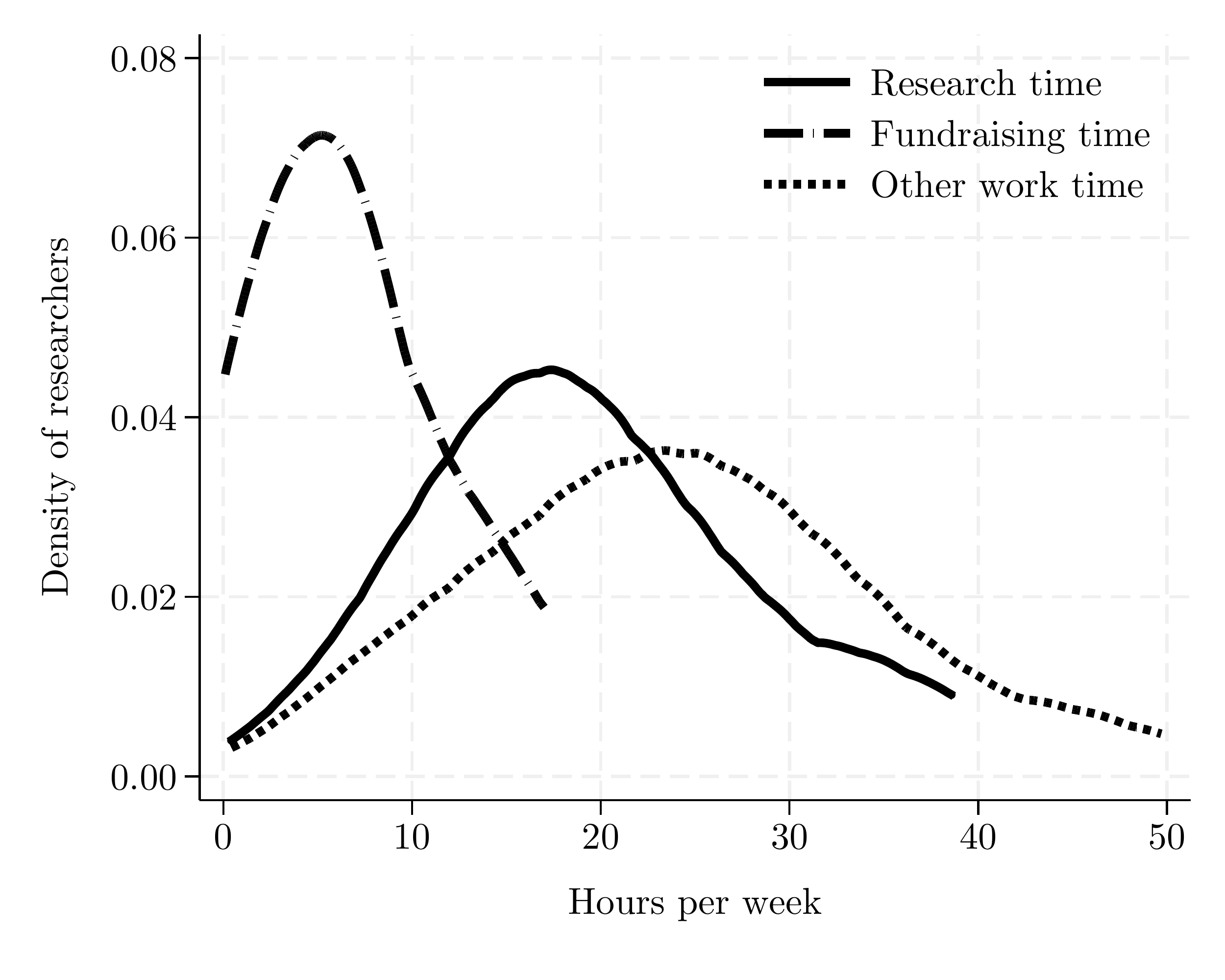}}
\subfloat[Correlation with expected grant \$]{
\includegraphics[width=0.475\linewidth, trim = 0mm 0mm 0mm 0mm , clip]{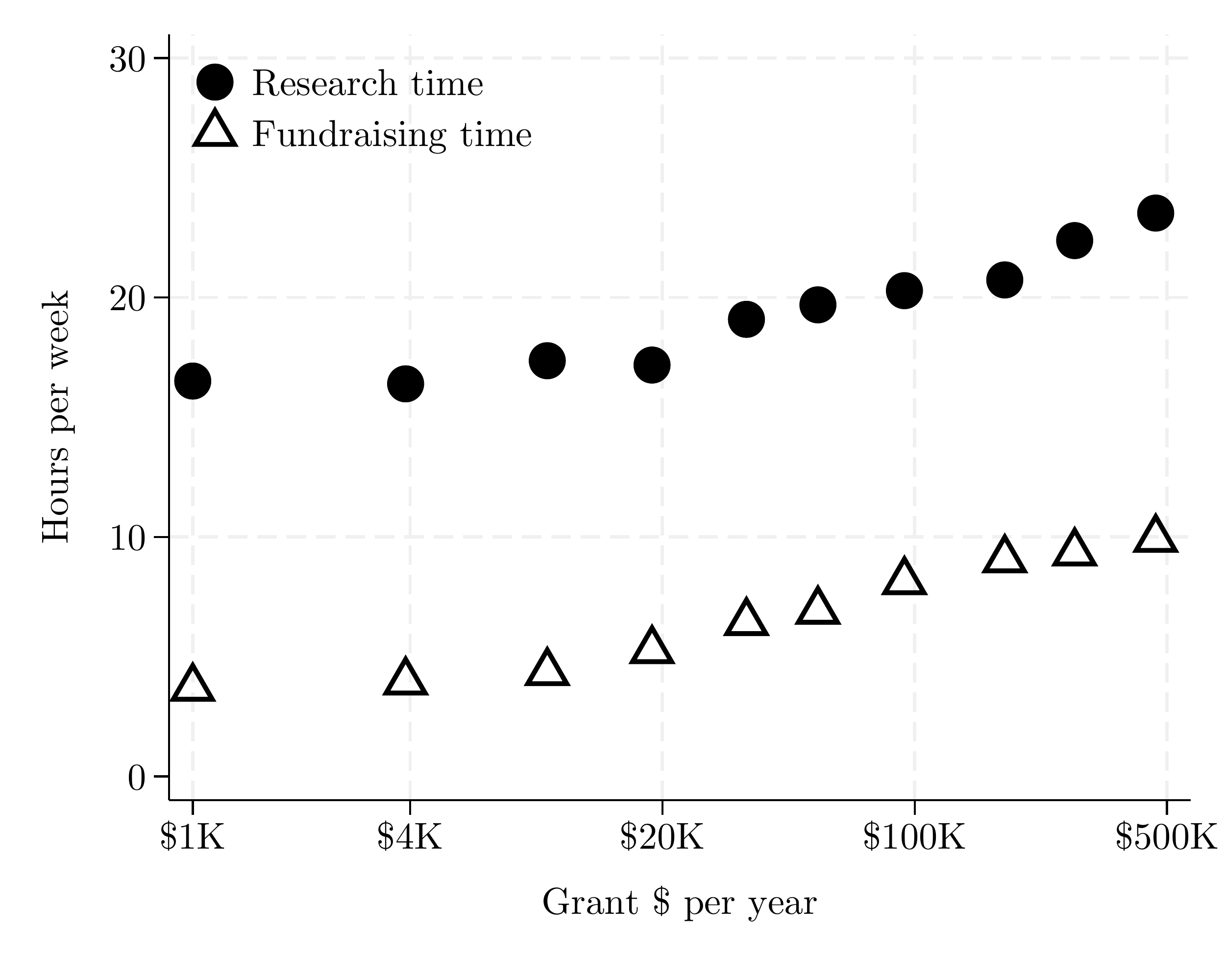}}
\begin{quote}\scriptsize
Note: Panel (a) shows the average amount of time spent on each activity. Panel (b) shows the distribution of average weekly time use that scientists forecast over the coming five years, where ``Other work'' includes teaching, administration, clinical, and any other work not reported as being either ``Research'' or ``Fundraising'' efforts. Panel (c) shows binned scatterplots of the unconditional relationship between scientists' forecasted fundraising expectations (i.e., how much grant funding they expect to obtain over the coming five years) and their time spent on research and fundraising; note the log scale of the $x$-axis.
\end{quote}
\end{figure}

Two components of the survey are relevant here. First, the survey asked professors to report their expected time allocations over the coming five years. The questions solicited professors' expected time they will spend on ``fundraising (e.g., writing grants)'' and ``research'' as well as other tasks (e.g., teaching, administration). It is important to emphasize that these time allocations are unverified forecasts. Thus measurement error is likely. However, it is plausible that this measurement error is common to all activities professors report (i.e., the error is not task-specific nor a function of the amount of time spent on the task). The bottom of Table \ref{tab_sumstat} reports the summary statistics for these time use measures, and Figure \ref{fig_sumstat} Panels (a--b) show further information on these distributions. 

Second, the survey asked professors how much grant funding they expect to obtain given the amount of time they expect to spend fundraising. These fundraising expectations (and professors pre-existing and/or guaranteed grant funding levels) are summarized at the bottom of Table \ref{tab_sumstat}. 

On average, professors in our sample expect to spend roughly 7 hours per week (15\% of total work hours) on fundraising, 20 hours per week (38\% of total work hours) on research, and 24 hours per week (47\% of total work hours) on all other tasks and duties. On average, professors expect that this 7 hours per week of fundraising efforts will translate to a total of \$130,000 in grant funding per year (over five years). Figure \ref{fig_sumstat} Panel (c) shows this clear positive relationship along with a clear positive relationship between research time and these funding expectations.

\subsection{Regression models and results}

Since the dependent variable, research time, is non-negative, I estimate a series of poisson regression models of the following form:
\begin{equation}\label{eq_main_log}
\mathbb{E}[ \text{Research}_i \,|\, \mathbf{W}_i; \bm{\uptheta}] = \exp\bigl(\alpha_2 + \beta_\text{F} \log(\text{Fundraising}_i) + \beta_\text{O} \log(\text{Other}_i) + g(\text{Grant \$}_i, \mathbf{X}_i) \bigr) \,,
\end{equation}
where $\beta_\text{F}$ is the focal parameter that identifies the elasticity of research time with respect to fundraising time. $\mathbf{W}_i$ is the vector of all observable variables (which contains $\mathbf{X}_i$, the vector of potential covariates to be selected by Lasso) and $\bm{\uptheta}$ is the vector of parameters. $g(\cdot)$ represents the control function. To flexibly control for grant funding, I use restricted cubic splines with four knots. When using Lasso, I include grant funding plus $\mathbf{X}_i$ and let Lasso select which controls to include in $g(\cdot)$.\footnote{All models are estimated either using the \texttt{poisson} or \texttt{ivpoisson gmm} commands in Stata{\tiny \copyright} depending on whether an instrumental variable is used or not. To incorporate Lasso, I first use the \texttt{popoisson} command to select a subset of the potential controls ($\mathbf{X}_i$) and then include all selected controls in the poisson regressions. I use all default settings for the Lasso-based \texttt{popoisson} command, and, in tables, report the number selected out of the number of potential controls. I use two sets of potential controls. The smaller set of potential controls is 57 variables spanning the socio-demographics, professional, and scientific features of each professor based on \cite{myers2023new}. The larger set is those same controls interacted with the grant funding controls less any perfectly collinear interactions.}

To provide a benchmark, I also test the null hypothesis that $\beta_\text{F}=\beta_\text{O}$. To the extent that other duties are as-good-as-randomly assigned (conditional on the controls), this provides a way of testing how large the effects of fundraising on research are relative to other tasks that affect scientists time allocations.\footnote{Though, as evidenced in Table \ref{tab_sumstat}, other tasks are on a different scale than fundraising (mean is almost three times higher). Thus, elasticities may be less relevant than linear relationships. For this and other reasons, I also report results from linear models in the Appendix.}

Table \ref{tab_main_log} reports the main regression results. First, Column (1) shows that research and fundraising time are unconditionally positively correlated. This may of course be representative of a true causal effect, but as noted previously, there is good reason to think that these two variable are positively correlated due to other unobservable forces. Columns (2--4) are all identified via conditional independence and include successively more control variables. When professors' other work time and grant funding expectations are included as controls, the association between fundraising and research time becomes smaller and statistically insignificant from zero (Col. 2). Including more control variables selected via Lasso does not change this (Cols. 3--4). In all of these initial cases I can reject the null that fundraising time affects research time in the same way that other work does. 

\begin{table}[htbp]\centering\footnotesize
\caption{Poisson regression results}
\label{tab_main_log}
{
\def\sym#1{\ifmmode^{#1}\else\(^{#1}\)\fi}
\begin{tabular}{l*{7}{c}}
\hline\hline
& \multicolumn{7}{c}{Research time} \\ \cline{2-8}
& \multicolumn{4}{c}{Poisson} & & \multicolumn{2}{c}{IV Poisson} \\ \cline{2-5} \cline{7-8}
                    &\multicolumn{1}{c}{(1)}         &\multicolumn{1}{c}{(2)}         &\multicolumn{1}{c}{(3)}         &\multicolumn{1}{c}{(4)}         &            &\multicolumn{1}{c}{(5)}         &\multicolumn{1}{c}{(6)}         \\
\hline 
                    &                     &                     &                     &                     &            &                     &                     \\
Fundraising time, log.&       0.090\sym{***}&       0.020         &       0.015         &       0.019         &            &      --0.077         &      --0.015         \\
                    &     (0.012)         &     (0.013)         &     (0.012)         &     (0.012)         &            &     (0.081)         &     (0.081)         \\
[0.5em]
Other work time, log.&                     &      --0.282\sym{***}&      --0.281\sym{***}&      --0.280\sym{***}&            &      --0.275\sym{***}&      --0.277\sym{***}\\
                    &                     &     (0.015)         &     (0.016)         &     (0.016)         &            &     (0.018)         &     (0.017)         \\
[0.5em]
\hline
$ p$--value($ \beta_{\text{F}}=\beta_{\text{O}}$)&                     &     $<$0.01         &     $<$0.01         &     $<$0.01         &            &        0.02         &     $<$0.01         \\
\hline  Incl. $ g(\scriptsize{\text{Grant \$}})$&                     &           Y         &           Y         &           Y         &            &           Y         &           Y         \\
Lasso $ g(\scriptsize{\text{Grant \$},X})$&                     &                     &           Y         &           Y         &            &           Y         &           Y         \\
$ X$ sel./poss.     &                     &                     &       19/57         &      24/219         &            &       19/57         &      24/219         \\
$ F$--stat.          &                     &                     &                     &                     &            &        33.1         &        33.4         \\
$ N$ obs.           &       2,640         &       2,640         &       2,640         &       2,640         &            &       2,640         &       2,640         \\
\hline\hline
\end{tabular}
}

\vspace{0.5em}
\begin{quote}\scriptsize
Note: Reports the results from estimating variations of Equation \ref{eq_main_log} using either poisson regression (Col. 1--4) or poisson regression with an instrumental variable (Col. 5--7, where the competition proxy is the instrumental variable). $p$--value($ \beta_{\text{F}}=\beta_{\text{O}}$) is based on a test of the null hypothesis that the coefficients on fundraising and other work time are equal. ``Incl. $g$(Grant \$)'' indicates that the flexible controls for expected grant funding are included. ``Lasso $g$(Grant \$,$X$)'' indicates that lasso was used to select from the flexible controls for expected grant funding as well as the additional set of covariates, and ``$X$ sel./poss.'' reports the number of selected and possible controls. Robust standard errors in parentheses; $^{*} p<0.1, ^{**} p<0.05, ^{***} p<0.01$.
\end{quote}
\end{table}

It is possible that there are still unobservable confounders driving the null effects identified in Columns (1--4) of Table \ref{tab_main_log}, so Columns (5--6) report the results using the competition proxy as an instrumental variable. The first stages of these models (reported in the Appendix Table \ref{tab_firststage_log}) indicate the instrument is relevant and that scientists in fields where their fundraising time is less productive (i.e., converts into less grant funding) spend less time fundraising. But when using an instrumental variable, the results are very similar. The point estimates are negative, but continue to be close to zero and are not statistically significant. Furthermore, I can still reject the null of equality between fundraising and other work time.

In the Appendix, I estimate linear versions of these same regressions and obtain similar estimates (Table \ref{tab_main_lin}). I find some negative effects of fundraising on research in the models with just controls, but when using the instrumental variable I again no longer reject the null hypothesis.

\subsection{Implications}

The results indicate that, on the margin, an hour less of fundraising effort does not change scientists' research effort. Why would fundraising time not detract from research time in the same way that other duties do? Let's consider the following representation of a scientists' objective function: 
\begin{equation}
\max_{F,R} p\bigl(G(F),F,R\bigr) - c(F,R,O) \,,
\end{equation}
where $p(\cdot)$ and $c(\cdot)$ are the production and cost functions, respectively, and all variables are the same as before. Scientists must choose how to allocate their fundraising and research time given the positive returns via $p$ and costs via $c$.

First, it seems very reasonable to assume that tasks are complements in the cost function (i.e., $\frac{\partial^2 c }{ \partial F \partial R} > 0$, etc.). Scientists all face the same constraint of 24 hours in a day, and as time is spent on one activity, the cost of spending additional time on other activities increases. Assuming this is true, then more time spent on a task that only enters the cost function should decrease time spent on other tasks. This is what I find in the relationship between research ($R$) and other tasks ($O$). However, if a task enters both the cost and production function, this need not be the case. If two tasks are complements in both the cost and production functions, then it is possible for a null relationship between the two to exist---more time on one task both increases the costs \emph{and} benefits of time on the other task, leading to no change. This is what I find in the relationship between research $R$ and fundraising $F$. And that this null result persists after conditioning on $G$ (which accounts for $\frac{\partial p}{ \partial G(F)} > 0$) suggests $F$ itself directly enters the production function.

This is consistent with the survey results of \cite{schneider2018results}, where scientists report that 38\% of fundraising effort involves meaningful science. If this magnitude is taken seriously, the null effect I estimate implies that one hour less of fundraising time is 23 minutes less of scientific work at the mean values of time allocations. Ignoring statistical significance, the most negative elasticities identified are in the range of --0.07, which implies 23 minutes less of ``fundraising'' based scientific work would spark approximately 12 minutes more of ``research'' based scientific work. Is the 37(=60--23) less minutes of non-scientific fundraising effort worth 11(=23--12) less minutes of total scientific effort? The answer to this tradeoff depends on the private and social costs and benefits of this effort and how it is influenced by competition, which I explore in the next section.

\section{Theory: Contest productivity and effort externalities}\label{sec_theory}

The prior section provides evidence that fundraising efforts do not affect research efforts, which is consistent with fundraising involving some inherently productive scientific tasks. How does the presence of this factor moderate the relationship between a contests' aggregate productivity (i.e., total benefits less total costs) and the level of competition (i.e., the share of potential applicants that receive an award)? In the following section, I answer this question using model of grant contests. In the model, there can be value generated by fundraising effort as an externality from the perspective of the scientist.\footnote{This reflects two points: (i) the positive externalities of all scientific activities are very large relative to their private costs, and (ii) while this entire paper is agnostic about ``when'' effort occurs (i.e., the models are static), it is reasonable to assume that scientific effort exerted in a funding contest will occur before it would have otherwise and society benefits from having the value of the scientists' work sooner (even if the same amount of value is generated).
However, there may also be negative externalities associated with application effort. Most notably, the social opportunity costs of this effort may be non-trivial because effort spent towards applications reduces the amount of effort that could be spent directly on other scientific activities. There also may be social costs of effort insofar as the delay in time necessary to accommodate the effort induced by the contest leads the social value of the grant to be reduced (e.g., in an emergency).}

In Section \ref{subsec_simple1} I outline an elementary model of contests with no effort externalities and contrast a traditional competitive model to a non-competitive lottery (where there is no competition). I evaluate which mechanism maximizes the planners' surplus, which I term the aggregate productivity. The lottery is more productive. In Section \ref{subsec_simple2}, I introduce positive effort externalities and show that the contest is more productive.  In Section \ref{subsec_expanded}, I build on the work of \cite{gross2019contest}, who in turn build on \cite{moldovanu2001optimal} and \cite{hoppe2009theory}, to show that this role of effort externalities is not unique to the highly stylized model in Sections \ref{subsec_simple1}--\ref{subsec_simple2}, but is also present in more realistic models of competitive grant contests.

\subsection{Simple model---Baseline}\label{subsec_simple1}

There is a single funder who will award a grant valued $v$ to one of $n$ scientists. Once awarded, the grant generates social benefits of $mv$, where $m>1$ determines the degree to which the grant generates positive externalities.\footnote{The private return to winning the grant is $v$ and the size of the positive externalities is $mv-v$.}
Scientists are equal in all regards (an assumption relaxed below) and can compete for the grant by submitting an application. Let $x_i$ be the effort scientist $i$ invests in their application with $cx_i$ denoting the disutility from this effort ($c>0$). I focus on the symmetric equilibrium where all scientists choose the same effort, $x^*$.

The funder's optimization problem is to choose a funding regime that maximizes the difference between the value generated by the grant and the total disutility of scientists' effort, $mv-ncx^*$.

One option is to use a ``contest'' regime where the probability that a scientist wins the grant is given by $\frac{x^*}{n x^*}$. It is straightforward to show the symmetric equilibrium gives $x^*=v\frac{(n-1)}{cn^2}$, which implies that the social value created is $(m - \frac{(n-1)}{n})v \equiv V_{\text{contest}}$.

Alternatively, the social planner could use a ``lottery'' regime by randomly allocating the grant to one scientist. Assuming this randomization is costless, the social value created is simply $mv \equiv V_{\text{lottery}}$. This will always generate more social value than in the first case ($V_{\text{lottery}}>V_{\text{contest}}$). This sort of argument either explicitly or implicitly underlies many of the proposals for lottery-based funding mechanisms: the introduction of randomness reduces the effort scientists invest in the grant competition. 

However, this argument ignores the reality that the effort put forward in grant competitions generates spillovers -- scientists can generate, refine, and share their ideas, regardless of whether or not they ultimately receive the funding. In other words, the effort exerted in these contest can involve positive (or negative) externalities.

\subsection{Simple model---With effort externalities}\label{subsec_simple2}

To see how effort externalities change the optimal regime choice, let $w$ determine the degree of these externalities so that $wx^*$ is the social benefit (or cost if $w<0$) each scientist's effort generates. This leads to a new problem for the funder: maximize $mv-n(c-w)x^*$. Scientists' strategies remain unchanged since, by definition, they do not take externalities into account when making their decisions.

The social value generated by the contest regime in this setting is $(m - (c-w)\frac{(n-1)}{cn}))v \equiv V'_{\text{contest}}$, while the value of the lottery regime is unchanged. The contest is more efficient if the social value of the effort externalities is greater than the private cost of that effort ($V'_{\text{contest}}>V_{\text{lottery}}$ if $w>c$). Furthermore, the social return to increasing competition ($n$) is increasing in the size of these externalities ($\partial V'_{\text{contest}}/\partial n \partial w > 0$). Similarly, larger competitions have more to gain from increasing the size of the effort externalities.

\subsection{Expanded model}\label{subsec_expanded}
The following closely matches \citeauthor{gross2019contest}'s (\citeyear{gross2019contest}) model. Some minor changes are also made to more closely align the model with traditional economic definitions of private versus social costs and benefits.

There is a single funding contest run by the social planner, whose budget is such that they can fund a proportion of scientists given by $p$. Each scientist $i$ has a single idea they need funding to pursue and the quality of that idea is given by the scalar $v_i \geq 0$, which is private information known only to $i$, but it is drawn from a publicly known distribution $F$.

The sequence of events and nature of the contest is as follows:
\begin{enumerate}
\item Scientists draw $v_i$ from $F$.
\item Scientists choose how much effort to exert to develop an application of quality $x_i$, where the private costs of applying are given by $c(v_i,x_i)$.
\item Applications are reviewed in a ``noisy'' process that imperfectly ranks applications per their quality ($x_i$) and applications are funded subject to the budget constraint given by $p$.
\item Scientists with funded applications receive a payoff of $v_i$. All scientists (funded or not) recoup some share of the private costs of applying, given by the parameter $k \in [0,1]$. The planner receives a payoff of $mv_i$ if $v_i$ is funded.\footnote{\cite{gross2019contest} also include a private payoff for funded applications (the $v_0$ parameter in their model) that does not enter into the planner's objective function. Since I am adopting the traditional economic definition of a social planner which incorporates the utility of all parties (scientists included), this term should also enter the planners' payoff. Instead, I am using the $m$ parameter to clearly generate a wedge between the private and social payoffs of the grant. This effectively sets $v_0$ to zero in terms of \citeauthor{gross2019contest}'s (\citeyear{gross2019contest}) model. Whether or not $v_0$ is included in my model turns out to be irrelevant since it is an additive term and unrelated to the degree of effort-based externalities, and the addition of the $m$ parameter only shifts the levels of the costs and benefits in the model which is not the focus.}
\end{enumerate}

Each scientists' optimal choice of application quality ($x^*_i$) is given by the bid function
\begin{equation}\label{eq_bidfunc}
b(v_i) = \argmax_{x_i} \; v_i \eta(x_i) - c(v_i,x_i)(1-k) \equiv x^*_i \;,
\end{equation}
where $\eta$ is the contest success function and describes the equilibrium funding probability for an application of quality $x_i$. This function depends on the payline ($p$), the distribution of realized applications, and the amount of noise in the review process.

The expected value of the contest for each applicant (if participating) is
\begin{equation}\label{eq_evapp}
v_i \eta\bigl(b(v_i)\bigr) - c \bigl(v_i,b(v_i)\bigr)(1-k)
\end{equation}
and zero otherwise. By assumption, only applicants with positive expected values will enter the contest. Also, unfunded applications generate no value beyond the private returns accrued to those applicants per the parameter $k$. 

In the case without any positive externalities from application effort, the expected value per award (the measure of the contests' aggregate productivity) for the social planner is
\begin{equation}\label{eq_evplan}
\frac{1}{p} \Biggl(\underbrace{\int mv \eta\bigl(b(v)\bigr)dF(v)}_{\text{benefit}} - \underbrace{\int c\bigl(v,b(v)\bigr)(1-k)dF(v)}_{\text{cost}}\Biggr) \;.
\end{equation} 

\cite{gross2019contest} show that the aggregate productivity of the contest is decreasing in the payline, $p$ (where the top $p$ applications are funded) as well as the value of instituting a lottery. is the same as the efficiency of a contest with a lottery line of $p$, even if the share that can be funded is less than $p$.

Formally, let $f$ be the share of applications that can be funded given the budget. In a standard contest, a payline $p$ is set to equal $f$ such that the top $p=f$ applications are funded. In the lottery mechanism, a lottery line $l$ is set greater than or equal to $f$ such that the top $l$ applications are entered into a randomized lottery where $f$ are funded. \cite{gross2019contest} show that the aggregate productivity of a payline of $p=f$ is the same as the productivity of a lottery line of $l\geq f$. For example, the productivity of a contest with a lottery line of 30\% is the same as the productivity of a non-lottery contest with a payline of 30\%, even if less than 30\% of the applications can be funded. Loosely speaking, the presence of a lottery lowers the optimal choice of effort for each scientist by reducing the effective level of competition they will face.

I incorporate effort externalities into this model by assuming that the planner's benefit now includes a ``bid-effort externality'' function $w(v,x)$, which can be thought of as the opposite of the private cost function, $c(v,x)$.\footnote{Very similar results (reported below) are obtained if one incorporates this externality by assuming that a portion of the idea's quality is realized and value by the social planner regardless of whether or not the idea is funded, with that portion realized being determined by the scientists' bid.} Now, the per-award benefit to the planner (omitting the cost component, which remains unchanged) is
\begin{equation}\label{eq_benext}
 \frac{1}{p} \int w(v,b(v)\bigr) + mv \eta\bigl(b(v)\bigr)dF(v) \;.
\end{equation}
The scientist's problem has not changed, so their bid function remains the same as in Equation \ref{eq_bidfunc}. 

Participants who work harder on their bid likely generate more externalities, but it is reasonable to assume there are decreasing returns (i.e., the initial conception of an idea may lead to many future ideas, but the 100th hour of work on a proposal is assumed to be spent on more wasteful activities). Thus, I choose a concave function for $w$ defined below.

The distributions and functional forms used are exactly as in \cite{gross2019contest}.\footnote{The distribution of idea quality is: $F(v)=1-(16/9)(1-v)^2)$, $v\in[0.25,1]$. The cost function and recovered costs are given by: $c(v,x)=x^2/v$ and $k=1/3$. The joint distribution of actual and evaluated quality are based on a Clayton copula with $\theta=10$.} In the baseline case, I set $m=1$, which eliminates the positive externalities generated by science in my model, but allows me to match \citeauthor{gross2019contest}'s (\citeyear{gross2019contest}) results exactly.

To parameterize the bid effort externalities, I choose a function form for $w$ that mirrors $c$ as a sort of inverse cost function:
\begin{equation}\label{eq_s}
w(v,x) = k v (x)^{1/r} \;,
\end{equation}
where $r$ influences the shape of the function. $w$ also includes the $k$ parameter that governs how much of their disutility from effort scientists recoup because that effort is spent on valuable research activities (per $kc(v,x)$).

\begin{figure}[htbp]\centering
\caption{Competition, externalities, and aggregate contest productivity}
\label{fig_theories}
\subfloat[Size of externality]{
\includegraphics[width=0.475\linewidth, trim = 0mm 0mm 0mm 0mm , clip]{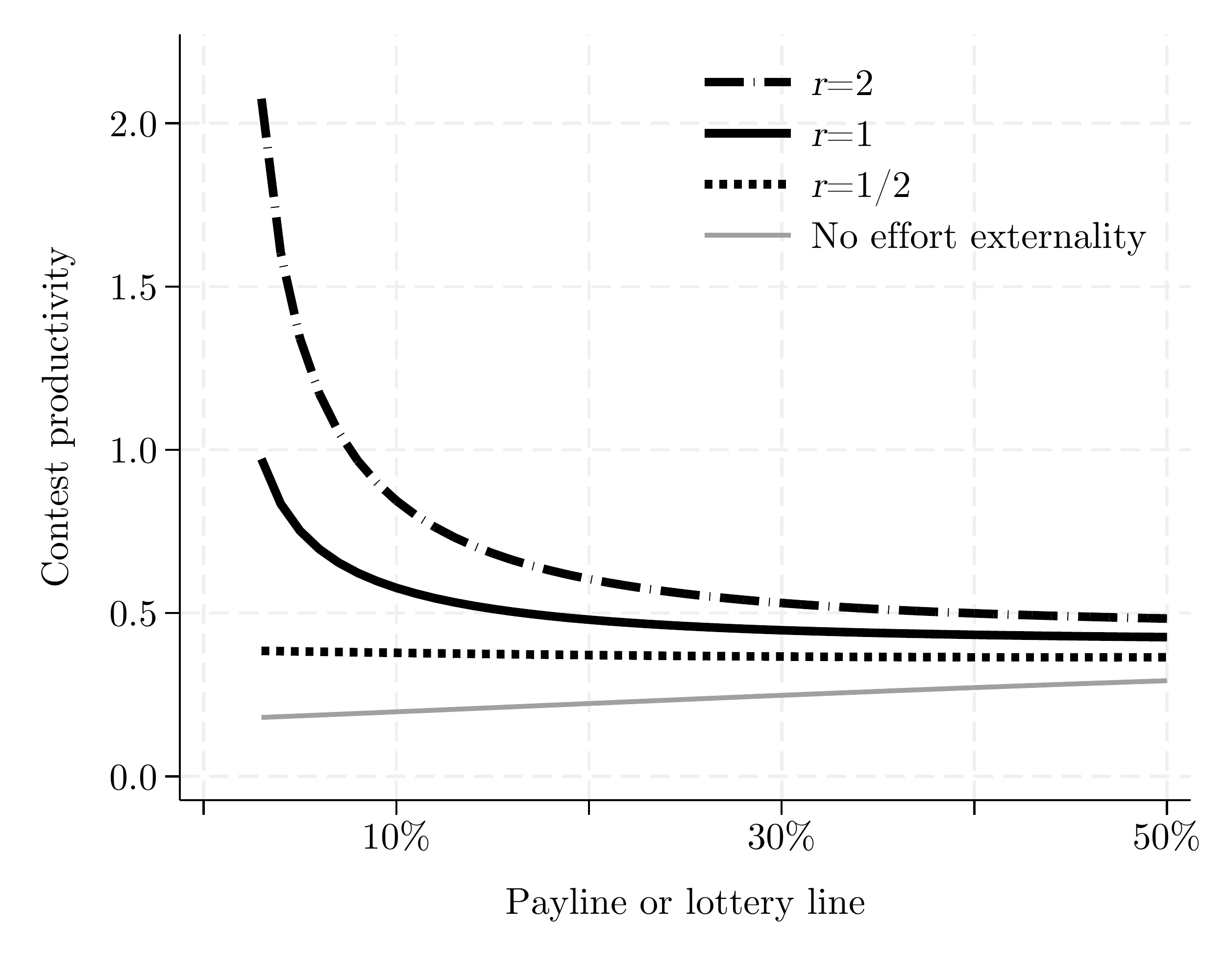}}
\subfloat[Nature of externality]{
\includegraphics[width=0.475\linewidth, trim = 0mm 0mm 0mm 0mm , clip]{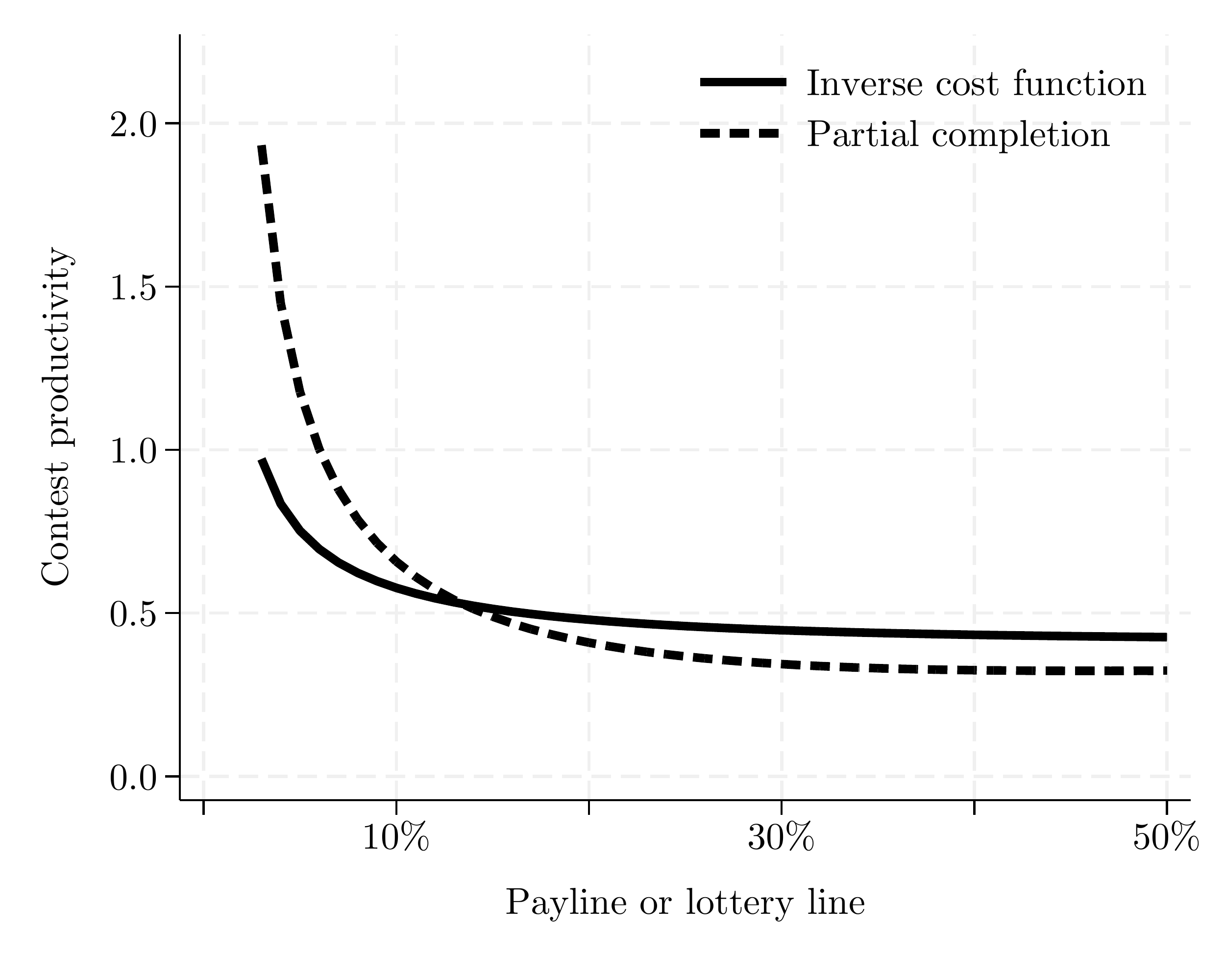}}
\begin{quote}\scriptsize
Note: Shows the productivity of a contest (total value per award) as a function of the contest payline (i.e., the percent of applications that are supported when ranked by quality). Panel (a) compares the baseline results of \cite{gross2019contest} (the ``No effort externality'' line) to scenarios where there are positive effort externalities (governed by the size of $r$). Panel (b) shows an alternative way of formulating the effort externality (as partial project completion) compared to the formulation shown in Panel (a) (as an inverse cost function).
\end{quote}
\end{figure}

Figure \ref{fig_theories} Panel (a) illustrates the contest's productivity under the baseline case of no effort externalities (exactly replicating \cite{gross2019contest}; see their Figure 3) as well as with externalities under a range of values for $r$. Because the support of the bid values lies below one, values of $r<1$, which dramatically reduce the size of the effort externalities, are necessary to eliminate the pattern whereby more competition increases the productivity of the contest. At $r=1/2$, the private costs and externalities perfectly offset, leading the productivity of the contest to be equal under all levels of competition. 

Another way one could incorporate bid effort externalities into this model would be to assume that a fraction of the idea's quality is produced as a function of each applicant's bid, regardless of funding status, with the remainder only being realized if the grant is awarded. For simplicity, let's also assume that the scientists' utility function is unchanged, which implies that they receive the full private value of the idea ($v_i$) only if funded whereas the social planner receives some portion of this value in the application process and (for those funded) the remainder. 

Formally, the the per-award benefit to the planner (omitting the cost component, which remains unchanged) is now
\begin{equation}\label{eq_benext_alt}
\frac{1}{p} \int w(b(v)) v + \Bigl(1-w(b(v)\bigr)\Bigr) v \eta(b(v))dF(v) \;,
\end{equation}
where $w(b(v)) v$ is the partial realization of all ideas due to application effort, and $\left(1-w\left(b\left(v\right)\right)\right) v$ is the realization of the remainder of each idea's value for the subset funded. Assuming the functional form of $w$ is again similar in nature to the other parts of the model, $w(b(v)) = b(v_i)^{(1/2)} / 3$, yields very similar results to the original formulation of externalities---see Figure \ref{fig_theories} Panel (b).

Overall, the theories illustrate that the aggregate productivity of a contest can vary dramatically depending on the nature of effort externalities. The more competitive the contest, the more there is to be gained from increasing the size of positive externalities. In other words, ensuring that scientists exert socially valuable effort to obtain grants is increasingly important as grant contests become more competitive.

\section{Discussion}\label{sec_discuss}

It is a waste when allocation mechanisms induce participants to exert effort that neither improves the allocative efficiency (i.e., the identification of scientists with higher quality ideas) nor the aggregate productivity of the mechanism (i.e., the total benefits less costs); this is a tautology. The question is \emph{to what degree} might competitive pressures of a contest improve efficiency or productivity. In the case of research grant contests, this paper is focused with the latter possibility---fundraising effort may be inherently valuable. If this is the case, then debating policies that influence competition in grant contests (e.g., the use of lotteries, the amount of preliminary results required) will involve debates about the tradeoffs of different kinds of efforts.

Using survey data on research-active professors who pursue grants, I find that the way they report their time use is consistent with fundraising generating value \emph{besides being a means to grant funding}. After accounting for many confounding factors, I find that more fundraising time neither spurs nor steals research time; I cannot reject the null hypothesis of no effect. But, the estimates are precise enough for me to reject the null hypothesis that fundraising time steals as much research time as professors' other duties (e.g., teaching and admin); the opportunity costs of fundraising (conditional on grant funding) appear much smaller than the opportunity costs of professors' other activities. An important caveat is that these time allocations are self-reported forecasts of averages over the next five years. On one hand, this introduces the possibility of significant measurement error and bias that may be correlated with professors' actual time allocations. On the other hand, analyzing these more aggregated patterns is informative of the longer-run equilibrium and is, in a sense, asking scientists how they plan to ``vote with their feet.'' While I can introduce a large number of potential control variables and also identify a plausibly exogenous source of fundraising time driven by competition across fields, there is still much room for improvement on both the measurement and identification of these variables and relationships.

Motivated by this result, I then show in theory how the presence of effort externalities (i.e., fundraising effort generates value the scientist does not capture) relates to the aggregate productivity of a grant contest. I build on \cite{gross2019contest} to show that when positive effort externalities exist, the aggregate productivity of the contest can be increasing in the level competition. Broadly speaking, this result is not new---it is a well-known result in economic theory that the social value of competition depends on whether competitive efforts generate positive or negative externalities. But this is one of the first times these externalities have been studied in a contest model designed to reflect the research grant environment.

This paper does not take a stance on whether the application processes of current funding environments are socially optimal. Unnecessary requirements or overly burdensome standards may be pervasive. Rather, this paper emphasizes that there can be large returns to designing grant applications that incentivize scientists to conduct science before the funding is awarded. The specifics of how to best do this is an open question and involves some tradeoffs not studied in this paper. Intuitively, application-centric work should probably not require substantial funding since this could handicap scientists in resource-intensive fields and those yet to acquire much funding. This suggests the type of effort that applications should incentivize should be towards theoretical and conceptual tasks: formalizing research questions, describing research designs and plans of work, etc.

Still, continued theoretical and empirical research is necessary to better understand the parameters highlighted here, as well as many of the complicated and dynamic factors not addressed, especially those related to allocative efficiency. For example, the theoretical models in this paper effectively assume a perfect correlation between scientists' research productivity and their ability to compete in the grant contests---the best researcher is also the best grant-writer. The pervasiveness of discussions surrounding ``grantsmanship'' amongst scientists suggests this is very likely not the case (e.g., \citealt{sauer2018grantsmanship,botham2020biosciences}). Furthermore, it is not always clear that faster science is higher quality science (\citealt{hill2021race}), so the ``quality-adjusted'' level of scientific effort within grant competitions may not be comparable.

More generally, the results imply that program evaluations comparing the outcomes of marginally funded and non-funded scientists (e.g., \citealt{jacob2011impact,myers2020elasticity}) may be limited in their policy relevance. While these research designs can provide strong internal validity with respect to scientist-level effects, they cannot identify the possibly large effects a contest has on all who participate. Given recent evidence on the declining productivity of the scientific sector (\citealt{bloom2020ideas}), the growing knowledge burden at the frontier of science (\citealt{jones2009burden}), and the increasing division of labor between academia and corporate science (\citealt{arora2020changing}), improving our ability to identify and incentivize the ``best'' scientists, whatever that may mean, continues to be an important issue.

\clearpage
\singlespacing
\bibliographystyle{apalike}
\bibliography{bibliography.bib}

\clearpage
\appendix
\section{Appendix}\label{sec_appendix}
\setcounter{figure}{0}
\renewcommand{\thefigure}{A\arabic{figure}}
\setcounter{table}{0}
\renewcommand{\thetable}{A\arabic{table}}
\setcounter{equation}{0}
\renewcommand{\theequation}{A\arabic{equation}}

\begin{table}[htbp]\centering\footnotesize
\caption{Distribution of narrow fields}
\label{tab_sumstat_dis}
{
\def\sym#1{\ifmmode^{#1}\else\(^{#1}\)\fi}
\begin{tabular}{l*{1}{rrrrr}}
\hline\hline
                    &       count&        mean&          sd\\
\hline
Agriculture         &       2,640&        0.03&        0.17\\
Biology             &       2,640&        0.06&        0.24\\
Business            &       2,640&        0.03&        0.16\\
Chemistry           &       2,640&        0.03&        0.17\\
Communication       &       2,640&        0.03&        0.16\\
Computer science    &       2,640&        0.02&        0.15\\
Economics           &       2,640&        0.02&        0.15\\
Education           &       2,640&        0.04&        0.19\\
Engineering         &       2,640&        0.08&        0.28\\
Geography           &       2,640&        0.06&        0.24\\
Humanities          &       2,640&        0.08&        0.28\\
Law                 &       2,640&        0.01&        0.11\\
Mathematics         &       2,640&        0.04&        0.18\\
Medical school      &       2,640&        0.20&        0.40\\
Medicine and health &       2,640&        0.07&        0.25\\
Other social sciences&       2,640&        0.04&        0.18\\
Physics             &       2,640&        0.05&        0.21\\
Political science   &       2,640&        0.04&        0.18\\
Psychology          &       2,640&        0.04&        0.21\\
Sociology           &       2,640&        0.03&        0.18\\
\hline\hline
\end{tabular}
}

\vspace{0.5em}
\begin{quote}\scriptsize
Note: Reports the share of the sample in each of the twenty narrower fields.
\end{quote}
\end{table}

\begin{table}[htbp]\centering\footnotesize
\caption{First stage, log. transformation}
\label{tab_firststage_log}
{
\def\sym#1{\ifmmode^{#1}\else\(^{#1}\)\fi}
\begin{tabular}{l*{5}{c}}
 \hline\hline
& \multicolumn{5}{c}{Fundraising time, log.} \\ \cline{2-6}
                    &\multicolumn{1}{c}{(1)}         &\multicolumn{1}{c}{(2)}         &\multicolumn{1}{c}{(3)}         &\multicolumn{1}{c}{(4)}         &\multicolumn{1}{c}{(5)}         \\
\hline \\
Funding intensity, std.&       0.202\sym{***}&       0.188\sym{***}&       0.060\sym{***}&       0.101\sym{***}&       0.100\sym{***}\\
                    &     (0.015)         &     (0.016)         &     (0.016)         &     (0.017)         &     (0.017)         \\
[0.5em]
Other work time, log.&                     &      --0.110\sym{***}&      --0.021         &       0.044         &       0.044         \\
                    &                     &     (0.032)         &     (0.027)         &     (0.029)         &     (0.029)         \\
[0.5em]
\hline
Incl. $ g(\scriptsize{\text{Grant \$}})$&                     &                     &           Y         &           Y         &           Y         \\
Lasso $ g(\scriptsize{\text{Grant \$},X})$&                     &                     &                     &           Y         &           Y         \\
$ X$ sel./poss.     &                     &                     &                     &           /         &           /         \\
$ R^2$              &        0.06         &        0.07         &        0.25         &        0.25         &        0.25         \\
$ N$ obs.           &       2,640         &       2,640         &       2,640         &       2,640         &       2,640         \\
\hline\hline
\end{tabular}
}

\vspace{0.5em}
\begin{quote}\scriptsize
Note: Reports the results from the first stage regressions of logged fundraising time on the focal instrumental variable, the proxy for grant funding competition (jackknife field-level average of grant funding expected per fundraising hour). $p$--value($ \beta_{\text{F}}=\beta_{\text{O}}$) is based on a test of the null hypothesis that the coefficients on fundraising and other work time are equal. ``Incl. $g$(Grant \$)'' indicates that the flexible controls for expected grant funding are included. ``Lasso $g$(Grant \$,$X$)'' indicates that lasso was used to select from the flexible controls for expected grant funding as well as the additional set of covariates, and ``$X$ sel./poss.'' reports the number of selected and possible controls. Robust standard errors in parentheses; $^{*} p<0.1, ^{**} p<0.05, ^{***} p<0.01$.
\end{quote}
\end{table}

\begin{table}[htbp]\centering\footnotesize
\caption{IV poisson regressions—Alternative instrumental variable formulations}
\label{tab_altz_log}
{
\def\sym#1{\ifmmode^{#1}\else\(^{#1}\)\fi}
\begin{tabular}{l*{4}{c}}
\hline\hline
& \multicolumn{4}{c}{Research time} \\ \cline{2-5}
                    &\multicolumn{1}{c}{(1)}         &\multicolumn{1}{c}{(2)}         &\multicolumn{1}{c}{(3)}         &\multicolumn{1}{c}{(4)}         \\
\hline
                    &                     &                     &                     &                     \\
Fundraising time, log.&      --0.015         &      --0.011         &      --0.033         &       0.027         \\
                    &     (0.081)         &     (0.081)         &     (0.047)         &     (0.096)         \\
[0.5em]
Other work time, log.&      --0.277\sym{***}&      --0.277\sym{***}&      --0.276\sym{***}&      --0.280\sym{***}\\
                    &     (0.017)         &     (0.017)         &     (0.017)         &     (0.018)         \\
[0.5em]
\hline
$ p$--value($ \beta_{\text{F}}=\beta_{\text{O}}$)&     $<$0.01         &     $<$0.01         &     $<$0.01         &     $<$0.01         \\
\hline IV version   &           1         &           2         &           3         &           4         \\
Incl. $ g(\scriptsize{\text{Grant \$}})$&           Y         &           Y         &           Y         &           Y         \\
Lasso $ g(\scriptsize{\text{Grant \$},X})$&           Y         &           Y         &           Y         &           Y         \\
$ X$ sel./poss.     &      24/219         &      24/219         &      24/219         &      24/219         \\
$ F$--stat.          &        33.4         &        34.2         &       112.5         &        23.2         \\
$ N$ obs.           &       2,640         &       2,640         &       2,640         &       2,640         \\
\hline\hline
\end{tabular}
}

\vspace{0.5em}
\begin{quote}\scriptsize
Note: Reports the results from using alternative instrumental variables for fundraising time. Column (1) uses the preferred instrument, jackknife field-level average of grant funding expected per fundraising hour and replicates Column (6) of Table \ref{tab_main_log}. The instrument for Column (2) is the jackknife field-level average of grant funding expected per share of time spent fundraising. The instrument for Column (3) is the jackknife field-level average of total funding (grant funding expected plus prior and guaranteed funding). The instrument for Column (4) is the jackknife field-level average of the share of total funding that is expected grant funding. All other elements are the same as in Table \ref{tab_main_log}. $p$--value($ \beta_{\text{F}}=\beta_{\text{O}}$) is based on a test of the null hypothesis that the coefficients on fundraising and other work time are equal. ``Incl. $g$(Grant \$)'' indicates that the flexible controls for expected grant funding are included. ``Lasso $g$(Grant \$,$X$)'' indicates that lasso was used to select from the flexible controls for expected grant funding as well as the additional set of covariates, and ``$X$ sel./poss.'' reports the number of selected and possible controls. Robust standard errors in parentheses; $^{*} p<0.1, ^{**} p<0.05, ^{***} p<0.01$.
\end{quote}
\end{table}

\begin{table}[htbp]\centering\footnotesize
\caption{Linear regression results}
\label{tab_main_lin}
{
\def\sym#1{\ifmmode^{#1}\else\(^{#1}\)\fi}
\begin{tabular}{l*{7}{c}}
\hline\hline
& \multicolumn{7}{c}{Research time} \\ \cline{2-8}
& \multicolumn{4}{c}{OLS} & & \multicolumn{2}{c}{2SLS} \\ \cline{2-5} \cline{7-8}
                    &\multicolumn{1}{c}{(1)}         &\multicolumn{1}{c}{(2)}         &\multicolumn{1}{c}{(3)}         &\multicolumn{1}{c}{(4)}         &            &\multicolumn{1}{c}{(5)}         &\multicolumn{1}{c}{(6)}         \\
\hline \\
Fundraising time    &       0.233\sym{***}&      --0.084\sym{**} &      --0.109\sym{***}&      --0.097\sym{**} &            &       0.009         &       0.070         \\
                    &     (0.035)         &     (0.038)         &     (0.039)         &     (0.039)         &            &     (0.291)         &     (0.296)         \\
[0.5em]
Other work time     &                     &      --0.374\sym{***}&      --0.342\sym{***}&      --0.341\sym{***}&            &      --0.340\sym{***}&      --0.337\sym{***}\\
                    &                     &     (0.016)         &     (0.017)         &     (0.017)         &            &     (0.017)         &     (0.017)         \\
[0.5em]
\hline
$ p$--value($ \beta_{\text{F}}=\beta_{\text{O}}$)&                     &     $<$0.01         &     $<$0.01         &     $<$0.01         &            &        0.22         &        0.16         \\
\hline  Incl. $ g(\scriptsize{\text{Grant \$}})$&                     &           Y         &           Y         &           Y         &            &           Y         &           Y         \\
Lasso $ g(\scriptsize{\text{Grant \$},X})$&                     &                     &           Y         &           Y         &            &           Y         &           Y         \\
$ X$ sel./poss.     &                     &                     &       18/57         &      22/219         &            &       18/57         &      22/219         \\
$ F$--stat.          &                     &                     &                     &                     &            &        42.6         &        41.1         \\
$ N$ obs.           &       2,640         &       2,640         &       2,640         &       2,640         &            &       2,640         &       2,640         \\
\hline\hline
\end{tabular}
}

\vspace{0.5em}
\begin{quote}\scriptsize
Note: Reports the results from estimating $\text{Research}_i  = \alpha + \beta_\text{F} \text{Fundraising}_i + \beta_\text{O} \text{Other}_i + g(\text{Grant \$}_i, \mathbf{X}_i) + \epsilon_{i}$ using either OLS (Col. 1--4) or 2SLS (Col. 5--7, where Funding intensity is the instrumental variable). $p$--value($ \beta_{\text{F}}=\beta_{\text{O}}$) is based on a test of the null hypothesis that the coefficients on fundraising and other work time are equal. ``Incl. $g$(Grant \$)'' indicates that the flexible controls for expected grant funding are included. ``Lasso $g$(Grant \$,$X$)'' indicates that lasso was used to select from the flexible controls for expected grant funding as well as the additional set of covariates, and ``$X$ sel./poss.'' reports the number of selected and possible controls. Robust standard errors in parentheses; $^{*} p<0.1, ^{**} p<0.05, ^{***} p<0.01$.
\end{quote}
\end{table}

\end{document}